\documentclass[twocolumn,preprintnumbers,aps,amsmath,amssymb,a4paper]{revtex4}

\usepackage{graphicx}
\usepackage{dcolumn}
\usepackage{color}

\begin{document}

\preprint{The following article has been published in Physics of Plasmas \textbf{16}, 072503 (2009). It can be found at \textcolor{blue}{ http://pop.aip.org/}.}

\title{Radial propagation of geodesic acoustic modes}

\author{Robert Hager}
 \email{robert.hager@ipp.mpg.de}

\author{Klaus Hallatschek}%
\affiliation{%
Max-Planck-Institut f\"ur Plasmaphysik\\
Boltzmannstra{\ss}e 2, D-85748, Garching, Germany
}%

\date{\today}


\begin{abstract}
The GAM group velocity is estimated from the ratio of the radial free energy flux to the total free energy applying gyrokinetic and two-fluid theory. This method is much more robust than approaches that calculate the group velocity directly and can be generalized to include additional physics, e.g. magnetic geometry. The results are verified with the gyrokinetic code GYRO [J. Candy and R. E. Waltz, J. Comp. Phys. \textbf{186}, pp. 545-581 (2003)], the two-fluid code NLET [K. Hallatschek and A. Zeiler, Physics of Plasmas \textbf{7}, pp. 2554-2564 (2000)], and analytical calculations. GAM propagation must be kept in mind when discussing the windows of GAM activity observed experimentally and the match between linear theory and experimental GAM frequencies.
\end{abstract}


\maketitle


\section{\label{sec:intro}Introduction}
Geodesic acoustic modes (GAMs) are axisymmetric poloidal $E \times B$ flows with finite frequencies coupled to up-down-antisymmetric pressure perturbations, which provide the restoring force of the oscillation.
Their frequency is proportional to $\sqrt{2} c_s/R$ with the sound speed defined as $c_s \equiv \sqrt{(T_{i0}+T_{e0})/m_i}$. This can be shown analytically \cite{winsor_gamfreq} and has been confirmed in recent measurements in ASDEX Upgrade \cite{conway_gamfreq} and DIII-D \cite{mckee_gamfreq}.
As GAMs are driven by the plasma turbulence, they are believed to play an important role in limiting the turbulence strength and thus might give rise to stable equilibria with reduced energy transport \cite{hall_transport}. GAMs have attracted widespread attention, covering for example the generation and damping of zonal flows \cite{zonca_freq,xu_gamdamp_prl,sugama_damp}, the GAM spectrum (for circular flux surfaces) \cite{zonca_freq} and GAM eigenmodes \cite{gao_itoh_eigenmode}.
In some works outward propagating GAMs are reported but no general rules determining the direction and speed of GAM propagation have been given \cite{zonca_freq,xu_gamdamp_prl,gao_itoh_eigenmode}, a crucial point concerning the radial windows of GAM activity observed in ASDEX Upgrade \cite{conway_gamamp} or DIII-D \cite{mckee_gamfreq} and the match between linear theory and experimental GAM frequencies \cite{hall_3dflow}.

The energy contained in a wave packet is transported with its group velocity as shown for example in Ref. \cite{swanson}.
Thus, one can calculate the group velocity of a GAM wave packet by compairing its total energy to its energy flux (Poynting flux). This provides a rather powerful tool for determining the direction and speed of GAM propagation.
It also allows a more general estimate of the maximal group velocity than a direct calculation of the dispersion relation, which requires much more effort and is restricted to a few simple cases.

The basic concepts of the method are demonstrated for the two-fluid equations for cold ions and large safety factor $q$. The generalization to warm ions, arbitrary safety factors and a gyrokinetic model is straightforward. The calculations are corroborated with exact analytical calculations for simple test cases and with numerical results obtained by the gyrokinetic code GYRO \cite{gyro} and the two-fluid code NLET \cite{hall_nlet} for various values of $q$ and $\tau=T_i / T_e$.

The structure of this article is as follows. In Sec. \ref{sec:bc} the equations for the free energy, its flux and the group velocity are derived within the two-fluid framework for $\tau=0$, $q \rightarrow \infty$, low $\beta$, and high aspect ratio circular magnetic geometry. The calculation is generalized in Secs. \ref{sec:general_fluid} and \ref{sec:general_kinetic} to the warm ion, finite $q$ case and to the gyrokinetic model. In Sec. \ref{sec:geometry}, the effects of plasma shaping on GAM propagation are discussed. Finally, the results are summarized in Sec. \ref{sec:discussion}.
%
%
\section{\label{sec:bc}Fluid Model for cold ions and infinite safety factor}
The units are chosen such that the magnetic drift velocity is unity. Density, $n$, temperature, $T_i$ and $T_e$, and electric potential perturbations $\phi$ are normalized to
\begin{equation}\label{eqn:basicu}
 \rho^\star n_0,\, \rho^\star T_{0,i/e},\, \rho^\star \frac{T_{0,e}}{e},
\end{equation}
respectively, where the subscript $0$ indicates the corresponding background value and $\rho^\star$ is given by $\rho_{se} / R$ with the major torus radius $R$, $c_{se}\equiv(T_{0,e}/m_i)^{1/2}$, and $\rho_{se}\equiv\left( m_i c_{se} \right)/\left(e B \right)$. The time scale is $t_0\equiv R/(2 c_{se})$.

\begin{figure*}
 \includegraphics[bb=14 14 519 187]{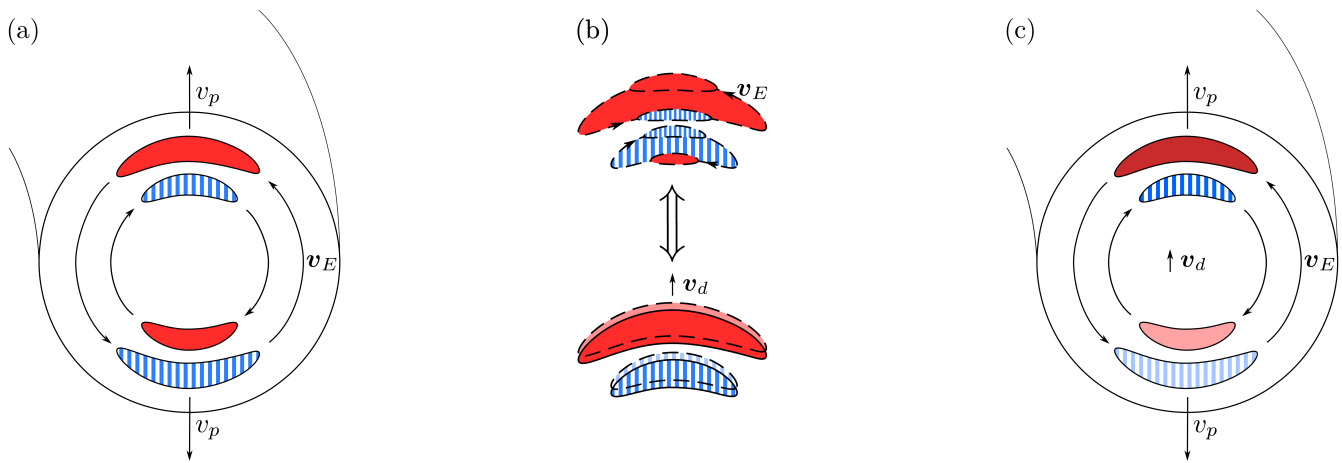}
 \caption{\label{fig:gam_cartoon}Sketch of a geodesic acoustic mode. (a) The poloidal $E\times B$-flow, moving with the GAM phase velocity $v_p$, leads to compression or expansion of the plasma (indicated by the filled and striped areas, respectively). Thus, an up-down antisymmetric $m=1$ density perturbation arises, which is phase-delayed against the flow by $\pi/2$. (b) Density perturbations are associated with an $E\times B$ flow, which leads to compression or expansion of the plasma, causing a drift of the density perturbations drifting with the ion magnetic inhomogeneity drifts. (c) Parallel drift and phase velocity enhance the density perturbations caused by the GAM poloidal rotation, antiparallel $v_d$ and $v_p$ lower the perturbations.}
\end{figure*}

Beforehand, it is useful to recall the main characteristics of the GAM. An $(m,n)=(0,0)$ $E \times B$ plasma flow, is generated by the flux-surface averaged electric field $-\nabla \phi_0$, with $\phi_0 \equiv \langle \phi \rangle$ and $\langle \dots \rangle$ indicating flux surface averaging.
The divergence of the flow due to the magnetic inhomogeneities gives rise to a mainly up-down-antisymmetric $m=1$ pressure perturbation [Fig. \ref{fig:gam_cartoon} (a)].
Since the compression of plasma requires work taken from the kinetic energy $(\nabla \phi_0)^2/2$ of the flow, a restoring force is generated, the flow is slowed down, stopped, and eventually reversed.
An oscillation between pressure perturbations and flow results, in which the maximal energy stored in the pressure perturbations is comparable to the initial kinetic energy.
\subsection{\label{subsec:bc_pflux}GAM Poynting flux and group velocity}
As a first approach, we calculate the Poynting flux (and group velocity) of the GAM for the cold ion two-fluid equations, neglecting sound waves ($q\rightarrow \infty$).
Since in the absence of perturbations the free energy is minimal, it is second order in the fluctuations. In the chosen framework, the free energy functional \cite{hall_erg}
is given by
\begin{equation}\label{eqn:edens_fluid}
  \left\langle E \right\rangle = \left\langle E_e + E_i \right\rangle = \left\langle \frac{n^2}{2} + \frac{\left( \nabla \phi_0 \right)^2}{2} \right\rangle,
\end{equation}
where $E_{e}$ and $E_{i}$ are the electron and the ion free energy density, respectively, $n^2/2$ is the energy of the electron density perturbations, and $(\nabla \phi_0)^2/2$ the ion kinetic energy.

The ion density fluctuations obey
\begin{equation}\label{eqn:fluid_neq_simple}
  \dot{n} - \Delta \dot{\phi} - \hat{C} \phi = 0,
\end{equation}
where $\hat{C} \equiv -\boldsymbol{v}_d \cdot \nabla$ and $\boldsymbol{v}_d \equiv -(1/2) (\hat{\boldsymbol{\kappa}} + R \nabla \ln B ) \times \hat{\boldsymbol{b}}$ is the sum of the curvature and $\nabla B$-drifts of the electron density fluctuations, and $\Delta \dot{\phi}$ is the divergence of the polarization current.
The electrons are assumed to be adiabatic
\begin{equation}\label{eqn:adiabatic}
 n=\phi-\phi_0, \, \left\langle n \right\rangle=0,
\end{equation}
because the GAM frequency is much smaller than the electron bounce and transit frequencies.
By combining (\ref{eqn:edens_fluid}), (\ref{eqn:fluid_neq_simple}), (\ref{eqn:adiabatic}), and representing the time derivative of $\langle E \rangle$ as the divergence of a radial Poynting flux one obtains
\begin{equation}\label{eqn:pflux_simple}
 \partial_t \left\langle E \right\rangle = - \left\langle \nabla \cdot \boldsymbol{S} \right\rangle = \left\langle -\nabla \cdot \left( \frac{\boldsymbol{v}_d n^2}{2} \right) + \nabla \cdot \left(n \nabla\dot{n}\right) \right\rangle.
\end{equation}
The first term, $\boldsymbol{v}_d n^2/2$, represents the flow of the energy of electron pressure perturbations in ion magnetic drift direction.
Since the radial component of $\boldsymbol{v}_d$ is up-down antisymmetric (for symmetric flux surfaces), this energy flux has a non-vanishing flux-surface average only if $n^2$ has an up-down asymmetry.
That such an asymmetry exists and that the energy transport is in the ion drift direction can be shown as follows.

Because due to adiabaticity (\ref{eqn:adiabatic}) the pressure fluctuations shown in Fig. \ref{fig:gam_cartoon} (a) are connected to potential fluctuations, they are encircled by $E\times B$-flows as indicated in Fig. \ref{fig:gam_cartoon} (b). Similar to the poloidal flow, the vortices lead to compression or expansion of the plasma owing to the magnetic field variations. This effect is equivalent to the advection of the density perturbation by the ion curvature drift, computed with the electron temperature.
Due to resonance between GAM phase and magnetic drift velocity, the pressure perturbations are enhanced at poloidal angles where drift and phase velocity are parallel, whereas they are weakened where those velocities are antiparallel.
Therefore, an up-down asymmetry of the energy density arises, which leads to a net radial energy transport through the flux surface
parallel to the phase velocity.
Due to the asymmetry requirement, this flux somewhat resembles neoclassical density or temperature transport.

Owing to adiabaticity (\ref{eqn:adiabatic}), pressure and potential perturbations are equal, so that the gradient of the local density fluctuations causes an electric field, whose time dependence gives rise to a polarization current density $-\nabla\dot{n}$.
Since the temperature is normalized to $T_{0,e}$, the term $-n \nabla\dot{n}$ can be interpreted as \textit{hydraulic} energy flux $p_e \boldsymbol{j}_{pol}$ consisting of the electron pressure $T_{0,e} n$ and the polarization current density. We will refer to it in the following as the polarization energy flux.
As the $E\times B$-flow associated with the density perturbations is proportional to the density gradients, the polarization energy flux can be regarded as the energy flux $-v_p (\nabla n)^2$, counterintuitively with the reversed phase velocity $v_p$.

The requirement of up-down asymmetry of the free energy density $n^2/2$ makes the curvature flux comparable in size to $-n \nabla\dot{n}$, a polarization effect. Whether the radial group and phase velocities eventually are parallel or antiparallel depends on the relative size of those two fluxes.

Next, the free energy and the Poynting flux, Eqs. (\ref{eqn:edens_fluid}) and (\ref{eqn:pflux_simple}), are evaluated in Fourier space for a circular high aspect ratio magnetic geometry.
Recalling that the curvature operator $\hat{C}$ is up-down antisymmetric for circular flux surfaces, one obtains an estimate for the density perturbations by splitting the ion density equation (\ref{eqn:fluid_neq_simple}) into an up-down antisymmetric and a symmetric part with the corresponding densities $n_a$ and $n_s$. Since the kinetic energy of the flow and the energy of the density fluctuations are of the same order, $k_r^2 \phi_0 \sim n^2$, only terms up first order in $k_r$ are kept in the antisymmetric equation. The symmetric density fluctuations are of second order in $k_r$. Thus, with $\hat{C}=-\sin(\theta) \partial_r$, the density becomes
\begin{align}\label{eqn:fludens_simple}
 n_a &\approx \frac{k_r}{\omega} \sin\left(\theta\right) \phi_0, \notag \\
 n_s &\approx \left[\frac{1}{\omega^2} \sin\left(\theta\right)^2 - 1\right] k_r^2 \phi_0.
\end{align}
Due to electron adiabaticity the GAM frequency $\omega$ determined by $\langle n \rangle = \langle n_s \rangle = 0$ is given by $\omega=2^{-1/2}$. Inserting (\ref{eqn:fludens_simple}) into (\ref{eqn:edens_fluid}) and (\ref{eqn:pflux_simple}) one obtains for the radial group velocity
\begin{equation}\label{eqn:vgroup_simple}
 v_{g,r} = \frac{\left\langle S_r \right\rangle}{\left\langle E \right\rangle} \approx -\frac{k_r}{2 \sqrt{2}}.
\end{equation}
Since the ratio of curvature to polarization flux is $-1/2$, the total Poynting flux and the group velocity are antiparallel to the phase velocity for cold ions.

The free energy approach only requires knowledge of the up-down antisymmetric density fluctuation $n_a$ and its symmetric correction $n_s$. The electron adiabaticity condition for $n$ as given by Eq. (\ref{eqn:fludens_simple}) only yields the GAM frequency to $0$th order in $k_r$. Thus, higher order corrections to the density have to be computed to calculate the group velocity directly from the dispersion relation. Hence, an advantage of the free energy approach is that less information is necessary compared to a direct calculation of the GAM frequency.

\begin{figure}
 \includegraphics[bb=3 2 220 128]{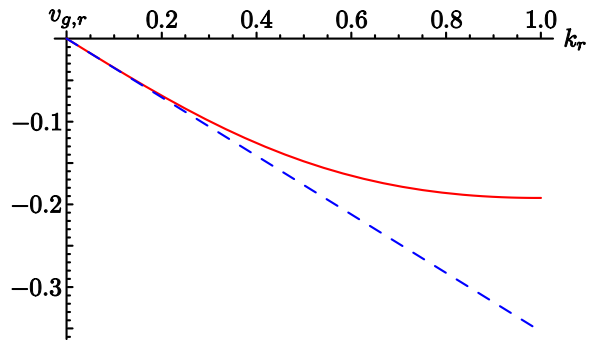}
 \caption{\label{fig:vgroup_ex}Exact GAM group velocity for $\tau=0$, $q \rightarrow \infty$ (\ref{eqn:vgroup_fluex}) (solid) and approximation (\ref{eqn:vgroup_simple}) (dashed).}
\end{figure}

To verify the approximation (\ref{eqn:vgroup_simple}), we give the exact solution of (\ref{eqn:fluid_neq_simple}),
\begin{equation}\label{eqn:fluid_ex1}
n=\phi_0 \left( \frac{\sqrt{\Omega^2-1}}{\sin \left(\theta\right) + \Omega} -1 \right), 
\end{equation}
with $\Omega=(\omega/k_r) (1+k_r^2)$.
The GAM frequency follows from the condition $\langle n \sin(\theta) \rangle=-\omega k_r \phi_0$, which is obtained from Eq. (\ref{eqn:fluid_neq_simple}) by using (\ref{eqn:adiabatic}), yielding $\omega=( 2 + k_r^2 )^{-1/2}$ and the corresponding radial group velocity
\begin{equation}\label{eqn:vgroup_fluex}
 v_{g,r} = -\frac{k_r}{\left(2+k_r^2\right)^{3/2}} \approx -\frac{k_r}{2 \sqrt{2}} + O\left[k_r^3\right],
\end{equation}
which to lowest order in $k_r$ is identical to the approximation (\ref{eqn:vgroup_simple}).
Figure \ref{fig:vgroup_ex} shows the exact group velocity in comparison with the approximated result. For small wavenumbers the latter converges against the exact result.
Deviations for larger wavenumbers are due to drift velocity resonances. The resonance condition can be obtained from Eq. (\ref{eqn:fluid_neq_simple}) giving
\begin{equation}\label{eqn:driftres}
 v_p \equiv \frac{\omega}{k_r} = \frac{\hat{\boldsymbol{k}}\cdot \boldsymbol{v}_d}{1+k^2} = -\frac{\sin \left(\theta\right)}{1+k_r^2}
\end{equation}
for circular flux surfaces.
A mode loses the character of a GAM, if, due to resonances, the energy of the density perturbations becomes significantly larger than the kinetic energy.
When the GAM frequency approaches a resonance,
the density amplitude becomes very large compared to the poloidal rotation
and dominates the mode.
The resonant pressure perturbations propagate with the group velocity of the resonant drift mode.
Accordingly, (\ref{eqn:fluid_neq_simple}) and (\ref{eqn:fludens_simple}) imply, that the modes discussed here are GAMs for small radial wavenumbers $k_r \ll 1$ only.
Therefore, the deviations of the approximate frequency from the exact one shown in Fig. \ref{fig:vgroup_ex} are due to the transition of the mode from a GAM to a magnetic drift mode.
\subsection{\label{subsec:bc_simulation}Numerical studies with NLET}
To corroborate the analytical insights, linear numerical studies were carried out with the two-fluid code NLET.
The computations were performed on a grid of 1024 radial and 32 parallel grid points with high aspect ratio circular geometry. The radial width of the computational domain was $400\,\rho_{se}$.
The remaining parameters are $\epsilon=a/R=0$, $\tau=0$, and $q$ ranging between $3$ and $30$.
Approximating $\epsilon=a/R=0$, $\tau=0$ and $q\rightarrow\infty$,
the GAMs have been initialized at time $t=0$ with an $(m,n)=(0,0)$ electrostatic potential, which then evolves selfconsistently.

\begin{figure}
 \includegraphics[bb=0 0 216 165]{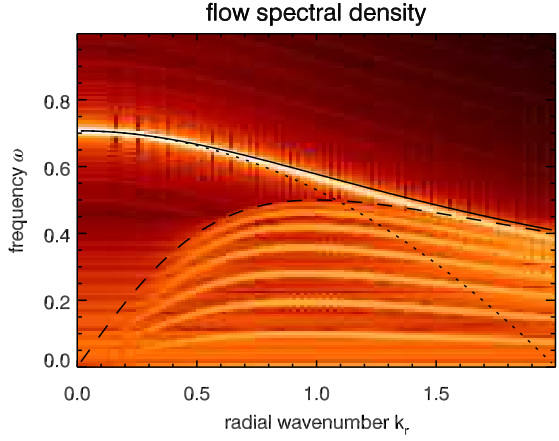}
 \caption{\label{fig:fluid_spec}NLET computed $\log$-color-coded GAM spectrum (Fourier transform of $\phi$) for $\tau=0$ and $q=30$ with exact analytical frequency (solid), approximate frequency [dotted, obtained by integrating Eq. (\ref{eqn:vgroup_simple})] and magnetic drift resonance frequency $k_r / \left(1 + k_r^2 \right)$ (dashed). The exact frequency (bright line) interpolates between the approximate and the resonance frequency. Also visible are additional drift modes enclosed by the resonance line.}
\end{figure}

The resulting spectral density of the radial $E\times B$-flow profile $v_E(r,t)$ is shown in Fig. \ref{fig:fluid_spec}.
The exact frequency, which also agrees with the numerical result, interpolates between the approximation obtained by integrating (\ref{eqn:vgroup_simple}) and the resonance frequency.
For $k_r\lesssim0.5$ the mode is obviously a GAM, whereas for larger $k_r$ it is gradually taken over by the magnetic drift resonance.
For $k_r \gtrsim 2$ the mode is completely dominated by resonant pressure perturbations as discussed in Sec. \ref{subsec:bc_pflux} and has lost the character of a GAM.
%
%
\section{\label{sec:general_fluid}Generalized fluid model}
The generalization of the preceding calculations is straightforward. Following Ref. \cite{hall_erg}, one can obtain the generalized ion free energy functional
\begin{multline}\label{eqn:fe_fluid}
 E_i=\frac{1}{2}\, \tau n^2 + \frac{3}{4}\, \tau T_{i}^2 + \frac{1}{2} \left[ \nabla \left( \phi + \tau \left( n + T_{i}\right)\right)\right]^2  \\
 + \frac{5}{4}\, \left(\nabla \tau T_{i}\right)^2 + \frac{1}{2} \left( v_{\parallel}^2 + \tau \left( \nabla v_{\parallel} \right)^2 \right),
\end{multline}
which includes ion temperature and parallel velocity. For warm ions, the ion density and temperature fluctuations contribute to the total free energy with $(1/2)\tau n^2+(3/4) \tau T_i^2$. The FLR (finite Larmor radius) heat flux contributes the energy density $(5/4)(\nabla \tau T_i)^2$. The energy density of the diamagnetic drift velocity increases by $\tau \nabla (n+T_i)$. For finite safety factor, the parallel flow $(1/2)v_\parallel^2$ and the FLR correction $(1/2) \tau ( \nabla v_{\parallel} )^2$ also contribute to the total energy density.
Fluid equations which exactly conserve the free energy functional \cite{hall_erg} are given by
\begin{eqnarray}\label{eqn:fluid_neq}
 \dot{n} &-& \Delta \left(\dot{\phi}+\tau \dot{n}+\tau \dot{T_i}\right) + \left( 1-\frac{5}{3} \tau \Delta \right) \partial_\parallel v_\parallel \notag \\
 &-& \hat{C} \left( \phi + \tau n + \tau T_i - \tau \Delta a \right) = 0, \\
 \label{eqn:fluid_Teq}
 \dot{T_i} &-& \frac{2}{3} \left[  \Delta \left(\dot{\phi}+\tau \dot{n}+\frac{7}{2}\tau \dot{T_i}\right) + \left( 1-\frac{7}{3} \tau \Delta \right) \partial_\parallel v_\parallel \right. \notag \\
 &-& \left.  \hat{C} \left( \phi + \tau n + \frac{7}{2}\tau T_i - \tau \Delta b \right)\right] = 0, \\
 \label{eqn:fluid_veq}
 \dot{v_{\parallel}} &-& \tau \Delta \dot{v_\parallel} + \left( 1-\frac{5}{3} \tau \Delta \right) \partial_\parallel \left( \phi + \tau \left(n+T_i\right) \right) \notag \\
 &-& \frac{5}{3} \tau^2 \Delta \partial_\parallel T_i - 2 \tau \hat{C} \left( \left( 1-\tau \Delta \right) v_\parallel \right) = 0,
\end{eqnarray}
with $a \equiv \alpha (\phi + \tau n)+\beta \tau T_i$ and $b\equiv \beta (\phi + \tau n) + \gamma \tau T_i$. For collisionless plasma, the three coefficients $\alpha$, $\beta$ and $\gamma$ are given by $(11/6,11/3,85/12)$.
Inserting (\ref{eqn:fluid_neq}-\ref{eqn:fluid_veq}) into $\partial_t E$ and writing the result in terms of divergences, we obtain the Poynting flux
\begin{align}\label{eqn:pflux_fluid}
 \partial_t \left\langle  E \right\rangle &= \left\langle -\nabla \cdot \left[\boldsymbol{v}_d \left( \frac{1}{2} \left( n+\tau p_i \right)^2 + \frac{5}{4} \left(\tau T_i\right)^2 + \tau v_\parallel^2 \right) \right] \right. \notag \\
 &+ \nabla \cdot \left[\left( n+\tau p_i \right) \nabla \left(\dot{n}+ \tau \dot{p_i}\right) +  \tau T_i \nabla \dot{\phi_0} \right. \notag\\
 &+ \left. \frac{5}{2} \tau^2 T_i \nabla \dot{T_i} +\tau v_\parallel \nabla \dot{v_\parallel} \right] - \tau \nabla \cdot \left\lbrace \boldsymbol{v}_d \left[ \frac{\Delta}{2} \left( \alpha c^2  \right. \right. \right. \notag \\
 &+ \left. 2\beta c d + \gamma d^2 \right) - \frac{3}{2} \left( \alpha \left(\nabla c \right)^2 + 2 \beta \nabla c \nabla d \right. \notag \\
 &+ \left. \left. \left. \left. \gamma \left(\nabla d \right)^2 \right) - \phi_0 \Delta a + \tau \left(  \Delta v_\parallel^2 -3\left( \nabla v_\parallel \right)^2 \right) \right] \right\rbrace \right\rangle,
\end{align}
in which $c \equiv \phi+\tau n$ and $d\equiv \tau T_i$.
The first two divergences on the right hand side of Eq. (\ref{eqn:pflux_fluid}) represent the advection of the fluctuation energy by the magnetic drifts and the polarization drift in complete analogy to the first term on the right hand side of Eq. (\ref{eqn:pflux_simple}). The last divergence in (\ref{eqn:pflux_fluid}) is an FLR correction to the first one.

\begin{figure}
 \includegraphics[bb=0 0 216 133]{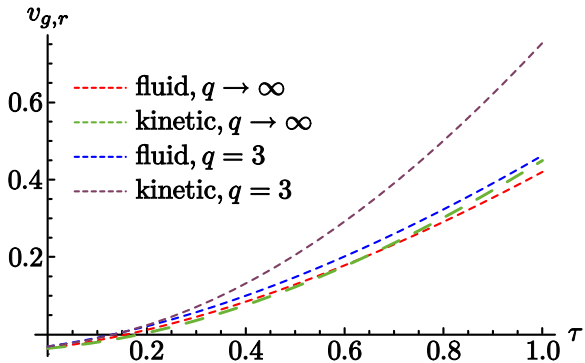}
 \caption{Ion temperature dependence of the fluid [Eq. (\ref{eqn:vgroup_fluid_app})] and the kinetic [Eq. (\ref{eqn:vgroup_gyro_app})] GAM group velocities for $k_r=0.1$.}
 \label{fig:vgroup_tau}
\end{figure}

We can approximate the two functionals (\ref{eqn:fe_fluid}), (\ref{eqn:pflux_fluid}), and the group velocity by splitting the fluctuations in (\ref{eqn:fluid_neq}-\ref{eqn:fluid_veq}) according to their up-down symmetry and keeping only the lowest order terms as in Sec. \ref{subsec:bc_pflux}.
When the GAM frequency approaches the sound frequency, the ratio of the energy densities of parallel flow velocity and density perturbations to the ion kinetic energy densities increases and tends to infinity close to resonance. Hence, the mode loses the character of a GAM.
Sound wave resonance is negligible, if $q \gg 1$ (in practice $q \gtrsim 3$ is sufficient).
Inserting the approximate perturbations into (\ref{eqn:fe_fluid}) and (\ref{eqn:pflux_fluid}), one obtains the group velocity
\begin{eqnarray}\label{eqn:vgroup_fluid_app}
 v_{g,r} &=& \frac{k_r}{2 \sqrt{6} \left( 3 + 5 \tau \right)^{3/2}} \left[ \left(-9 +21 \tau + 189 \tau^2 + 265 \tau^3 \right) \right. \notag \\
 &+& \frac{1}{4 q^2 \left( 3 + 5 \tau \right)} \left( 135 + 1026 \tau + 3324 \tau^2 \right.  \notag \\
 &+& \left. \left. +5360 \tau^3 + 3675 \tau^4  \right) \right].
\end{eqnarray}
(For details see appendix \ref{app:poynting_fluxes}.)
All additional terms compared to Eq. (\ref{eqn:vgroup_simple}) are positive, which causes the group velocity to change sign at $\tau\approx 0.16$ (Fig. \ref{fig:vgroup_tau}).
When calculating GAM eigenmodes as in Ref. \cite{gao_itoh_eigenmode}, regions of evanescent and propagating GAM would be switched at this critical $\tau$ because the group velocity is reversed. Existence and properties of global GAM eigenmodes might be relevant for the efficiency of GAM excitation in the same way as they are for Alfv\'{e}n wave excitation [e.g., for toroidal Alfv\'{e}n eigenmodes (TAE)]. Since GAMs have recently been found to play an important role in nonlinear turbulence saturation \cite{waltz_gamtransp}, their propagation might influence turbulent transport.
%
%
%
\section{\label{sec:general_kinetic}Gyrokinetic model}
Generalizing the previous discussion to gyrokinetic theory we use the linear model \cite{frieman_gkeq}
\begin{equation}\label{eqn:gkeq}
 \partial_t f + \frac{\boldsymbol{v}_\mathrm{d}}{\tau} \cdot \nabla_\perp \left( \tau f + F_0 \hat{J}_0 \phi \right) + \frac{v_\parallel}{\tau} \partial_\parallel \left( \tau f + F_0 \hat{J}_0 \phi \right) = 0
\end{equation}
with the quasineutrality condition
\begin{equation}\label{eqn:quasineutrality}
 n + \frac{1-\hat{\Gamma}_0}{\tau}\, \phi - \int \hat{J}_0 f\,\mathrm{d}^3 v = 0.
\end{equation}
The velocity $\boldsymbol{v}_{\mathrm{d}}$ is the sum of the curvature and $\nabla B$ drift of the individual particles. $F_0$ is the thermal background distribution function, which is normalized such that $\int F_0\,\mathrm{d}^3v=1$.
Gyro-averaging is represented by the operator $\hat{J}_0$ and the thermal average of $\hat{J}_0^2$ is defined by $\hat{\Gamma}_0 \equiv \int F_0 \hat{J}_0^2\, \mathrm{d}^3 v$. The Fourier representations of $\hat{J}_0$ and $\hat{J}_0^2$ are $J_0 (\tau^{1/2} v_\perp k_r)$ and $\Gamma_0 (k_r) \equiv \exp(-\tau k_r^2) I_0 (\tau k_r^2)$, respectively, with the Bessel function of the first kind $J_0$ and the modified Bessel function of the first kind $I_0$.
The ion free energy density \cite{hall_erg} is
\begin{equation}\label{eqn:fe_gyro}
 E_i=\int \frac{1}{\tau} \frac{\left(\tau f \right)^2}{2 F_0} \mathrm{d}^3 v + \frac{1}{2}\, \phi \frac{1-\hat{\Gamma}_0}{\tau} \phi,
\end{equation}
in which first term represents the energy of the fluctuations of the gyro-averaged distribution function, and the second one the energy of the gyrophase dependent fluctuations, i.e. the plasma polarization.
Using (\ref{eqn:gkeq}), $\partial_t E$ can be written as
\begin{align}\label{eqn:pflux_gyro}
 \partial_t \left\langle E \right\rangle &= \left\langle -\int \left[ \nabla \cdot \frac{\boldsymbol{v}_d}{\tau} \frac{K^2}{2 F_0} - \left\lbrace \frac{\boldsymbol{v}_d}{\tau} \cdot \nabla K , n \right\rbrace_{J_0} \right. \right. \notag \\
 &- \left\lbrace \frac{\boldsymbol{v}_\parallel}{\tau} \cdot \nabla \tau f , n  \right\rbrace_{J_0} + \left\lbrace \tau f, \frac{\boldsymbol{v}_d}{\tau} \cdot \nabla \phi_0 \right\rbrace_{J_0} \notag \\
 &+ \left. \left\lbrace n F_0, \frac{\boldsymbol{v}_d}{\tau} \cdot \nabla \phi_0 \right\rbrace_{J_0^2} \right] \mathrm{d}^3 v -\frac{1}{2} \left\lbrace \phi,\dot{\phi} \right\rbrace_{\frac{1-\Gamma_0}{\tau}} \notag \\
 &+ \left. \nabla \cdot \left( \phi_0 \hat{\chi} \partial_t \boldsymbol{E}_0 \right) \right\rangle
\end{align}
with $K \equiv \tau f + J_0 n F_0$. The susceptibility operator $\hat{\chi}$ is defined by its Fourier transform $\chi(k_r) \equiv (1-\Gamma_0 (k_r))/(\tau k_r^2)$.
The brackets denote
\begin{equation}\label{eqn:pbracket}
 \left\lbrace a,b \right\rbrace_{K}  \equiv a \left( K \ast b \right) - b \left( K \ast a\right),
\end{equation}
with $\ast$ indicating convolutions. As shown in appendix \ref{app:comm_div}, $\left\lbrace a,b \right\rbrace_{K}$ can always be written as divergence, provided the kernel $K$ is symmetric.

The first term on the right hand side of Eq. (\ref{eqn:pflux_gyro}) represents the advection of the free energy of gyro-averaged fluctuations by magnetic drifts.
The remaining four terms in the integral are FLR corrections (e.g., gyroviscosity and the Bakshi-Linsker effect) equivalent to the FLR energy fluxes in Eq. (\ref{eqn:pflux_fluid}).
The last two terms describe the polarization energy flux. They can be expressed as
\begin{equation}\label{eqn:polflux_gyro}
-\chi n \left( \omega k_r n \right) - v_p\, \frac{\chi \left\vert k_r \phi \right\vert^2}{2}\, \frac{\partial \ln \chi}{\partial \ln k_r},
\end{equation}
where the first term is the gyrokinetic equivalent to the fluid term $-n \nabla \dot{n}$, and the second one is an FLR correction.
Keeping only the lowest order terms, we compute an approximation of the group velocity as in Secs. \ref{subsec:bc_pflux} and \ref{sec:general_fluid} by splitting $f$ and $n$ in (\ref{eqn:fe_gyro}) and (\ref{eqn:pflux_gyro}) according to their up-down symmetry, and even and odd terms in $v_\parallel$.
One obtains the radial group velocity (App. \ref{app:poynting_fluxes})
\begin{eqnarray}\label{eqn:vgroup_gyro_app}
 v_\mathrm{g} &=& \frac{k_r}{8 \sqrt{2} \left( 4 + 7 \tau \right)^{3/2}} \left[ \left(-32 +24 \tau+ 586\tau^2+1277\tau^3 \right) \right. \notag \\
 &+&  \frac{1}{4 q^2 \left( 4 + 7 \tau \right)^{2}}\, \left( 640 +8736 \tau+ 55000\tau^2 \right. \notag \\
 &+& \left. \left. 215268 \tau^3 + 560074 \tau^4 + 526209 \tau^5 \right) \right].
\end{eqnarray}

Far from drift and sound resonances, for $k_r \ll 1$, $q \gg 1$, the gyrokinetic equation can be solved alternatively by a power series expansion in terms of $k_r$ and $q$ yielding an identical result, also in agreement with \cite{zonca_freq}.
The difference between the kinetic (\ref{eqn:vgroup_gyro_app}) and the fluid (\ref{eqn:vgroup_fluid_app}) group velocity is negligible for $q\rightarrow \infty$ (Fig. \ref{fig:vgroup_tau}) and for $\tau=0$. For $q \gtrsim 3$  and $\tau\gtrsim0.2$ the kinetic group velocity tends to be higher than the fluid one ($75\%$ at $\tau=1$, increasing with $\tau$). This is caused by an earlier onset of the coupling to parallel modes due to hot particles.

\begin{figure}
 \includegraphics[bb=0 0 216 165]{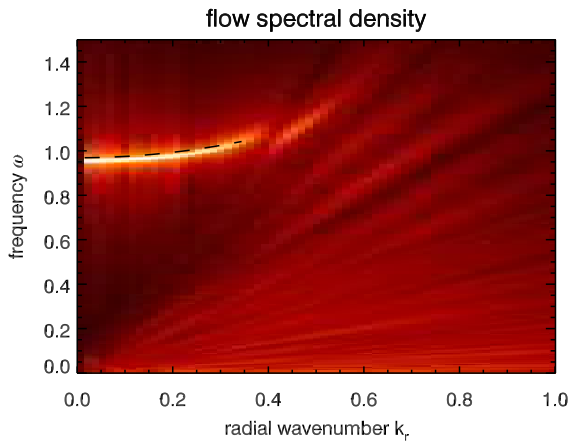}
 \caption{\label{fig:gyro_spec}GYRO computed log-color-coded GAM spectrum (Fourier transform of $\phi$) for $\tau=0.5$ and large safety factor $q=30$. The analytical kinetic frequency (\ref{eqn:vgroup_gyro_app}) is indicated by the dashed black line. Due to damping of the resonant modes, the simulated dispersion ends at $k_r \sim 0.4$, as for warm ions the number of resonant particles increases with $\tau$, so that Landau damping has to be taken into account.}
\end{figure}

Computational analyses with GYRO confirm the analytical results obtained from the kinetic and the fluid calculation. The simulations have been performed on a grid of 800 radial, 1 toroidal, 6 orbit and 8 energy gridpoints with a radial box size of $400 \rho_{se}$, $0\leqslant\tau\leqslant0.5$, and $3\leqslant q \leqslant 30$. Beforehand, agreement with NLET for cold ions has been checked.
An example for $\tau=0.5$ and $q=30$ is shown in Fig. \ref{fig:gyro_spec} together with the frequency obtained by integrating Eq. (\ref{eqn:vgroup_gyro_app}).
%
%
%
\section{\label{sec:geometry}Magnetic geometry effects}
Let us now turn to the effects of plasma shaping on the Poynting flux. Assuming, for simplicity, cold ions, infinite safety factor, and neglecting the polarization drift, one can approximate Eq. (\ref{eqn:fluid_neq_simple}) at $\theta = \pm\pi/2$
(where $\boldsymbol{k}_r$, $\boldsymbol{v}_d$, and $\boldsymbol{v}_p$ are parallel)
\begin{equation}\label{eqn:pressure_est}
 n \approx \frac{2 v_E}{\left(\omega-\boldsymbol{k}_r\cdot\boldsymbol{v}_d\right) R}=\frac{2 v_E}{R \boldsymbol{k}_r \cdot \left(\boldsymbol{v}_p-\boldsymbol{v}_d\right)},
\end{equation}
in which $R$ is the major radius, and $v_E=k_r \phi_0$ is the $E\times B$-drift velocity. For $k_r\ll1$, equivalent to $v_p \gg v_d$, Eq. (\ref{eqn:pressure_est}) implies
\begin{equation}\label{eqn:pressure_est_vd}
 n^2 \approx \frac{4 v_E^2}{\omega^2 R^2} \left( 1+ \frac{2 v_{d,r}}{v_p} \right),
\end{equation}
$v_{d,r}$ being the radial component of $\boldsymbol{v}_d$.
Accordingly, the flux surface averages of the two terms on the right hand side of Eq. (\ref{eqn:pflux_simple}) (the Poynting flux) can be estimated by
\begin{equation}\label{eqn:pfluxes_geo}
 \frac{1}{2} \left\langle v_{d,r} n^2 \right\rangle \approx \frac{v_d^2}{v_p} \left\langle n^2 \right\rangle = \frac{k_r v_d^2}{\omega} \left\langle n^2 \right\rangle,\, k_r \omega \left\langle n^2 \right\rangle \approx k_r \frac{2 v_E^2}{\omega R^2}.
\end{equation}
Since $\langle E \rangle \approx \langle n^2 \rangle$ and $v_d=1$ in the units defined in \ref{sec:bc}, the group velocity is of order $O[k_r]$. As the typical velocity scale of turbulent motion is $v_{dia}$, $v_{dia}\gg v_d$, and $k_r\ll1$, GAMs generally propagate much slower than turbulence and the magnetic drifts.

The specific magnetic geometry enters the calculation by means of the factors $k_r(\theta)/\omega$ and $k_r(\theta) \omega$ in the neoclassical and the polarization and FLR fluxes, respectively. Experimentally, only the radial wavenumber at the outboard midplane, $k_0$, is known.
However, the GAM pressure fluctuations, energy fluxes, and group velocity have to be estimated at $\theta=\pm \pi/2$, where $k_r$ is smaller due to, e.g., ellipticity or Shafranov shift.
For an elliptic Miller equilibrium \cite{miller}, the radial wavenumber at $\theta=\pi/2$ is given by
\begin{equation}\label{eqn:geo_correction_k}
 k_r = \frac{1+\partial_r R}{\kappa + r \partial_r \kappa} k_0,
\end{equation}
where $\kappa$ is the elongation and $r$ the minor radius at the outboard midplane ($r$ and $R$ refer to the flux surface center).

\begin{figure}
\includegraphics[bb=0 0 219 269]{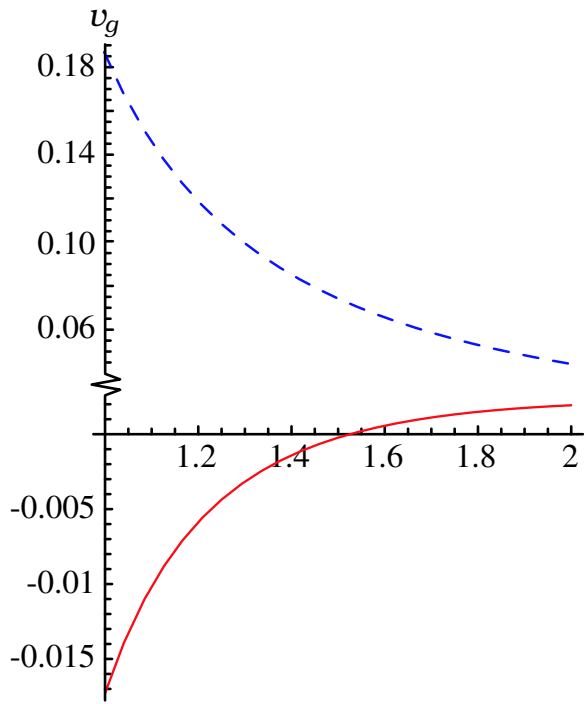}
 \caption{Estimates of the radial group velocities for $k_r=0.1$, $\tau=0$ (solid) and $\tau=1$ (dashed) plotted against the elongation $\kappa$.}
 \label{fig:vgroup_kappa}
\end{figure}

Typical values of the geometry parameters are \cite{miller} $\partial_r \kappa=(\kappa-1)/r$, $\partial_r R = -1/3$ and aspect ratio $A=3.5$.
The $\kappa$ dependence of $\omega$ for $k_r=0$ can be obtained numerically \cite{hall_3dflow} and is rather accurately described for $\kappa = 1..2$ and $q>1$ by
\begin{equation}\label{eqn:gamfreq_geo}
 \omega\left(\kappa\right) \approx \left(1+\frac{1}{4q^2}\right) \sqrt{1+\frac{5}{3}\tau} \frac{2.97}{1+3.73 \kappa}.
\end{equation}
Other possible parametrizations are discussed in Ref. \cite{conway_gamfreq}.
Substitution of these parameters into Eq. (\ref{eqn:geo_correction_k}) yields
\begin{equation}\label{eqn:geo_corrections}
 k_r \approx \frac{2}{3 \left( 2 \kappa - 1\right)} k_0.
\end{equation}
Multiplying the magnetic drift energy fluxes (\ref{eqn:curv1}-\ref{eqn:curv3}) with $k_r^2 \omega(1)/(k_0^2 \omega(\kappa))$ and the polarization energy fluxes (\ref{eqn:curvco}-\ref{eqn:pol4}) with $k_r^2 \omega(\kappa)/(k_0^2 \omega(1))$, and defining $\kappa\equiv 1+\delta\kappa$, we obtain an approximation of the group velocities at the outboard midplane for $\tau=0$
\begin{multline}\label{eqn:vgroup_geo_t0}
 v_{g,r} \left( \delta\kappa \right) \approx \frac{k_0}{0.32+2.15\delta\kappa+5.31\delta\kappa^2+5.54\delta\kappa^3+2\delta\kappa^4} \\ \left[ -0.050-0.021\delta\kappa+0.19\delta\kappa^2 +0.062\delta\kappa^3 \right. \\
 +\left. \frac{0.062+0.26\delta\kappa+0.33\delta\kappa^2+0.11\delta\kappa^3}{q^2}\right],
\end{multline}
and $\tau=1$
\begin{multline}\label{eqn:vgroup_geo_t1}
 v_{g,r} \left( \delta\kappa \right) \approx
 \frac{k_0}{0.32+2.15\delta\kappa+5.31\delta\kappa^2+5.54\delta\kappa^3+2\delta\kappa^4} \\
 \left[0.59+1.84\delta\kappa+1.57\delta\kappa^2 +0.52\delta\kappa^3 \right. \\
 +\left. \frac{0.54+2.51\delta\kappa+3.45\delta\kappa^2+1.14\delta\kappa^3}{q^2}\right].
\end{multline}
Figure \ref{fig:vgroup_kappa} shows the $\kappa$ dependence of the group velocity. In the cold ion case, increasing plasma elongation leads to a change of sign of the group velocity when the ``neoclassical'' fluxes become larger than the polarization terms.
Compared to circular flux surfaces, the group velocity for warm ions is reduced in elliptic geometry. Therefore, far from resonances, $v_g \sim k_r v_d$ remains small compared to the diamagnetic velocity and the magnetic drift. Our cold ion approximation agrees with numerical studies performed with NLET.

However, in single-null configuration near the separatrix, the perturbations vanish at the X-point due to the magnetic null and GAMs are located opposite towards the X-point. Consequently, the neoclassical energy flux is $\langle v_{d,r} n^2 \rangle/2 \sim v_d \langle n^2 \rangle$ and $v_g \sim v_d$, because the polarization energy flux is one order smaller and can be neglected. Hence, independent of the ion temperature, GAMs propagate in the ion magnetic drift direction, which is usually directed towards the X-point, i.e. radially inward. The GAM dispersion must be linear in this case.
In the up-down symmetric case, the radial mode structure can show standing waves while in up-down antisymmetric geometry propagating waves are expected as the wavenumbers of incoming and reflected wave at a cutoff layer are different.

%
%
%
\section{\label{sec:discussion}Summary and discussion}
We have given estimates for the radial group velocity of GAMs using an energy approach, which exploits Poynting's theorem.

In the first step, the GAM energy flux (\ref{eqn:pflux_simple}) has been calculated in a two-fluid framework for cold ions and infinite safety factor, and two kinds of transports have been identified. The ``neoclassical'' energy flux, which represents the advection of the free energy of the fluctuations with the magnetic inhomogeneity drift, $\langle v_{d,r} E_{fluc} \rangle$, is always parallel to the phase velocity. However, the polarization energy flux, $\langle -k_r^2 \rho_{se}^2 v_p E_{fluc} \rangle$,  is always antiparallel to the phase velocity. Therefore, the one of the two fluxes which is dominant controls the propagation direction of the GAMs.
This result has been generalized to include ion temperature and parallel flows (\ref{eqn:pflux_fluid}) and to a gyrokinetic model (\ref{eqn:pflux_gyro}), in which equivalent energy transports have been found.

In the second step, the group velocity, which is given by the ratio of Poynting flux to total free energy, has been evaluated for circular high aspect ratio flux surfaces [Eqs.(\ref{eqn:vgroup_fluid_app}), (\ref{eqn:vgroup_gyro_app})]
and estimated for elliptic plasmas [Eqs. (\ref{eqn:vgroup_geo_t0}), (\ref{eqn:vgroup_geo_t1})].
The results agree with NLET and GYRO computations, and analytical calculations. For cold ions, group and phase velocity have been found to be opposite. For $\tau \gtrsim 0.2$  or $\kappa \gtrsim 1.5$, they are parallel.
The neoclassical energy flux requires an up-down asymmetry of $E_{fluc}$ to give non-zero values. In case of up-down symmetric flux surfaces, the asymmetry is of order $k_r \rho_{se} E_{fluc}$, which makes the neoclassical comparable to the polarization energy flux. Therefore, the group velocity $v_{g,r}=\langle S_r \rangle/\langle E_{fluc} \rangle$ is of order $k_r \rho_{se} v_d$.

Due to resonances with magnetic drift modes (not drift waves) and sound waves, the propagation speed of GAMs is limited. Close to such a resonance, the poloidal rotation becomes negligible compared to the remaining degrees of freedom, as the amplitudes of density and temperature, or parallel flow velocity diverge. The characteristics of the resulting mode are completely determined by the resonant mode, and it should not be called GAM any longer.
The resonances restrict GAMs to $k_r \ll 1$ and $q \gg 1$ and limit the group velocity.

Since for single-null configuration, which is quite common in today's experiments, the fluctuations vanish at the X-point, the neoclassical energy flux is of order $v_d E_{fluc}$, and GAMs propagate at the magnetic drift velocity. More precisely, keeping only the first term on the right hand side of (\ref{eqn:pflux_fluid}), neglecting $v_\parallel$, and using $T=2n/3$, the ratio of Poynting flux to free energy yields the group velocities $v_{g,r} \approx v_d$ for $\tau=0$ and $v_{g,r} \approx 3 v_d$ for $\tau=1$. Since usually $v_d$ is directed towards the X-point, GAMs propagate radially inward in this case.
Overall, the GAM group velocity has been shown to be much smaller than the diamagnetic velocity, which is the typical scale of turbulent motion. In up-down symmetric magnetic geometries, $v_{g,r}$ is even much smaller than the magnetic inhomogeneity drift.

Calculating the group velocity by means of the Poynting flux is advantageous compared to a direct calculation of the GAM dispersion relation because only lowest order approximations of the fluctuations are required (Sec. \ref{subsec:bc_pflux}) to gain insights into effects induced by ion temperature, sound waves, or magnetic geometry. Requiring relatively small efforts, the energy approach enabled us to calculate the group velocity rather accurately for circular high aspect ratio geometry, to estimate it for elliptic Miller equilibria and to predict the effect of the X-point in single-null configuration. Since the formation and quality of the H-mode is dependent on whether the magnetic drift is directed towards or away from the X-point, one may speculate if there is a relation to GAM propagation.


\appendix
\section{\label{app:comm_div}Gyrokinetic commutators as divergences}
The brackets defined in Eq. (\ref{eqn:pbracket}) in Sec. \ref{sec:general_kinetic} can always be written as a divergence. Consider the expression
\begin{equation}\label{eqn:commutator}
 C = a \left( K \ast b \right) - b \left( K \ast a \right)
\end{equation}
where $a(x)$ and $b(x)$, $K\left(x\right)=K\left(-x\right)$ is a symmetric convolution kernel and $\ast$ indicates the convolution operation. Because of this symmetry one may express equation (\ref{eqn:commutator}) as
\begin{equation}\label{eqn:div_flux}
 \partial_z J \equiv C = \partial_z \int\int L\left(z-x,z-y\right) a\left(x\right) b\left(y\right) \mathrm{d}x \,\mathrm{d}y
\end{equation}
with $L\left(x,y\right)=\left[\theta\left(x\right)-\theta\left(y\right) \right] K\left(x-y\right)$.
For purely harmonic waves $a \left( x \right),\, b \left( x \right) = \Re \left\lbrace (A,B) \exp \left(i k x -i \omega t\right)\right\rbrace$, the resulting flux $J$ is given by
\begin{equation}
 J = \left(\partial_z\right)^{-1} \left\lbrace a,b \right\rbrace_{\hat{K}} = \Im\left\lbrace A^\ast B \right\rbrace \partial_k \hat{K}\left(k\right)
\end{equation}
where $\hat{K}$ is the Fourier transform of $K$.
The application of this theorem to, for instance, the first bracket in Eq. (\ref{eqn:pflux_gyro}) yields the corresponding free energy flux
\begin{equation}
 J_{\mathrm{FLR}} = \int J_1 \left( k_\perp \right) \frac{\mathbf{k} \cdot \mathbf{v}_\mathrm{d}}{\tau} \Re \left\lbrace n^\ast \left( \tau f + F_0 J_0 n\right) \right\rbrace \mathrm{d}^3 v.
\end{equation}
\section{\label{app:poynting_fluxes}Individual Poynting fluxes}
Evaluating Eq. (\ref{eqn:pflux_fluid}) for circular flux surfaces by using estimates of the up-down symmetric and antisymmetric fluctuation amplitudes (as described in Sec. \ref{subsec:bc_pflux}) the individual Poynting fluxes result up to order $O[k_r^3]$,
\begin{multline}
\label{eqn:curv1}
 \left\langle\frac{\boldsymbol{v}_d}{2} \left( n+\tau p_i \right)^2\right\rangle = \frac{k_r^3 \phi_0^2}{6 \sqrt{6 \left( 3 + 5 \tau \right)}} \left[9 + 30\tau + 55\tau^2 \right.\\
 + \left. \frac{1}{4 q^2} \left( 81 + 342\tau + 455 \tau^2 \right) \right],
\end{multline}
\begin{multline}
\label{eqn:curv2}
 \left\langle \frac{5 \boldsymbol{v}_d}{4} \left( \tau T_i \right)^2 \right\rangle = \frac{5 \tau^2 k_r^3 \phi_0^2}{3 \sqrt{6} \left( 3 + 5 \tau \right)^{3/2}}
 \left[3 + 20\tau  \right. \\
 +\left. \frac{\left( 81 + 387\tau + 500\tau^2 \right)}{4 q^2 \left( 3+5\tau \right)} \right],
\end{multline}
\begin{equation}
\label{eqn:curv3}
 \left\langle \boldsymbol{v}_d \tau v_\parallel^2 \right\rangle = \frac{\tau k_r^3 \phi_0^2 \left( 9 + 39\tau + 50\tau^2 \right)}{q^2 \sqrt{6} \left( 3 + 5 \tau \right)^{3/2}}
\end{equation}
\begin{multline}
\label{eqn:curvco}
 \left\langle \tau \boldsymbol{v}_d \left[ -\frac{\Delta}{2} \left( \alpha c^2 + 2\beta c d + \gamma d^2 \right) + \phi_0 \Delta a \right. \right. \\
 + \frac{3}{2} \left( \alpha \left(\nabla c \right)^2 + 2 \beta \nabla c \nabla d + \gamma \left(\nabla d \right)^2 \right) \\
 - \left. \left. \tau \left(  \Delta v_\parallel^2 -3\left( \nabla v_\parallel \right)^2 \right) \right] \right\rangle= \\
 = \left(1+\frac{1}{4q^2}\right) \frac{11 \tau k_r^3 \phi_0^2 \left(3+7\tau \right)}{3 \sqrt{6} \left( 3+5\tau \right)^{1/2}},
\end{multline}
\begin{multline}
\label{eqn:pol1}
 \left\langle -\left(n+\tau p_i \right) \nabla \left( \dot{n} +\tau \dot{p_i} \right) \right\rangle = \\
 =-\left(1+\frac{1}{4q^2}\right) \frac{k_r^3 \phi_0^2 \left(3+5\tau\right)^{3/2}}{3 \sqrt{6}},
\end{multline}
\begin{equation}
\label{eqn:pol2}
 \left\langle -\frac{5}{2} \tau^2 T_i \nabla \dot{T_i} \right\rangle = -\left(1+\frac{3}{4q^2} \right) \frac{10 \tau^2 k_r^3 \phi_0^2 }{3\sqrt{6} \left(3+5\tau \right)^{1/2}},
\end{equation}
\begin{equation}
\label{eqn:pol3}
 \left\langle -\tau T_i \nabla \dot{\phi_0} \right\rangle = -\left(1+\frac{1}{4q^2} \right) \frac{10 \tau^2 k_r^3 \phi_0^2 }{3\sqrt{6} \left(3+5\tau \right)^{1/2}},
\end{equation}
\begin{equation}
\label{eqn:pol4}
 \left\langle -\tau v_\parallel \nabla \dot{v}_\parallel \right\rangle = -\frac{\tau k_r^3 \phi_0^2 \left(3+5\tau \right)^{1/2}}{2 \sqrt{6} q^2}.
\end{equation}

For the gyrokinetic framework, Eq. (\ref{eqn:pflux_gyro}), the individual fluxes are
\begin{multline}
\label{eqn:S1a}
 \left\langle \int \frac{\boldsymbol{v}_d}{\tau} \frac{K^2}{2 F_0}\, \mathrm{d}^3v \right\rangle = \frac{k_r^3 \phi_0^2}{4\sqrt{2} \left(4+7\tau \right)^{3/2}} \left[ 16 + 140\tau \right. \\
 + \left. 481\tau^2 + 747\tau^3 + \frac{1}{q^2 \left( 4+7\tau \right)^2} \left( 576 \right. \right. \\
 + 6576 \tau + 36068\tau^2 + 130144\tau^3 \\
 + \left. \left. 317687\tau^4 + 293067 \tau^5 \right) \right],
\end{multline}
\begin{equation}
\label{eqn:S2a}
 \left\langle \nabla^{-1} \left(-\int \left\lbrace \frac{\boldsymbol{v}_d}{\tau} \cdot \nabla K , n \right\rbrace_{J_0} \mathrm{d}^3v \right) \right\rangle = 0,
\end{equation}
\begin{multline}
\label{eqn:Spara}
 \left\langle \nabla^{-1} \left(-\int \left\lbrace \frac{\boldsymbol{v}_\parallel}{\tau} \cdot \nabla \tau f , n  \right\rbrace_{J_0} \mathrm{d}^3v \right) \right\rangle =\\
 =-\frac{\sqrt{2} \tau k_r^3 \phi_0^2 \left(2+5\tau \right)}{q^2  \left( 4+7\tau \right)^{3/2}},
\end{multline}
\begin{multline}
\label{eqn:S3a1}
  \left\langle \nabla^{-1} \left(\int \left\lbrace \tau f, \frac{\boldsymbol{v}_d}{\tau} \cdot \nabla \phi_0 \right\rbrace_{J_0} \mathrm{d}^3v \right) \right\rangle = \\
 =-\frac{\tau^2 k_r^3 \phi_0^2}{2\sqrt{2} \left(4+7\tau \right)^{1/2}} \left[ 13 + \frac{28 + 196\tau + 163\tau^2}{q^2 \left( 4+7\tau \right)^2} \right],
\end{multline}
\begin{multline}
\label{eqn:S3a2}
  \left\langle \nabla^{-1} \left(\int \left\lbrace n F_0, \frac{\boldsymbol{v}_d}{\tau} \cdot \nabla \phi_0 \right\rbrace_{J_0^2}  \mathrm{d}^3v \right) \right\rangle = \\
 =-\frac{\tau k_r^3 \phi_0^2}{2\sqrt{2} \left(4+7\tau \right)^{1/2}} \left[ 12 + \frac{48 + 168\tau + 60\tau^2}{q^2 \left( 4+7\tau \right)^2} \right],
\end{multline}
\begin{multline}
\label{eqn:S4aS5a}
 \left\langle \nabla^{-1} \left( \frac{1}{2} \left\lbrace \phi,\dot{\phi} \right\rbrace_{\frac{1-\Gamma_0}{\tau}} \right) - \phi_0 \hat{\chi} \dot{\boldsymbol{E}}_0 \right\rangle =\\
 =\frac{k_e^3 \phi_0^2}{8 \sqrt{2} \left( 4+7\tau \right)^{1/2}} \left[ -16 + 12 \tau +21\tau^2 + \frac{1}{q^2 \left( 4+7\tau \right)^2} \right. \\
 \left. \left( -192 + 656\tau - 252\tau^2 + 612\tau^3 + 483\tau^4 \right) \right].
\end{multline}




\end{document}